\documentstyle[twocolumn,aps, epsfig]{revtex}

    \def\z{\noindent}  
 
    \def\sqr#1#2{{\vcenter{\vbox{\hrule height .#2pt
                             \hbox{\vrule width .#2pt height#1pt \kern#1pt
                                   \vrule width .#2pt}
                             \hrule height .#2pt}}}}

\begin{document}
\wideabs{ \title{Space charge limited flow of a thin electron beam 
confined by a strong magnetic field}

\author{A.Rokhlenko and J. L. Lebowitz\footnotemark }
\address{Department of Mathematics,
Rutgers University\\ Piscataway, NJ 08854-8019}

\maketitle
\begin{abstract}
            
An approximate analytic theory is developed and implemented
numerically for calculating the space charge limited current and
electric field of a thin cylindrical beam or current sheet between two
wide parallel electrodes.  The flow is confined by a sufficiently
strong magnetic field.  Assuming that the potential and current
density are almost homogeneous in the direction transversal to the
flow we compute the beam current and profile by a variational
method. We find that the average current density scales as the
reciprocal of the beam width when the latter becomes very small. The total
cylindrical beam current thus decreases proportionly to its diameter
while the total current of a sheet becomes almost independent of the
width in this regime.

\medskip

\z PACS: 52.27.Jt; 52.59.Sa; 52.59.Wd; 85.45.Bz

\end{abstract}}
   
\narrowtext
\noindent 1. {\it Introduction}

\smallskip
Space charge limited electron flow in two and three dimensions
presents a difficult nonlinear mathematical problem whose solution is 
important for many practical applications. In the design of 
high power electron beams [1-4] the one dimensional
Child-Langmuir limit (CLL) [5] has been a benchmark for almost a
century but corrections have to be made for the ``current wings'' near 
the boundaries of the flow. Consequently the current 
of narrow beams [6-8] show great divergences from CLL.

\footnotetext{*Also Department of Physics} 

In a previous article [9] we considered a planar emitting region
whose width $2a$ is much larger than the cathode-anode distance (1 in
our dimensionless units). In the case of narrow beams, $2a\ll 1$,
considered here, the moderate variations of the potential in the
transversal direction will be used to reduce the dimensionality of the
problem. We assume that the two parallel electrodes are 
large enough for the electric field to be homogeneous far
away from the beam. A magnetic field directed along the $y$-axis keeps
the current, which is emitted either by a long narrow straight strip or 
a circular disk, perpendicular to the electrodes without
spreading out as shown in Fig.\ 1.

The outline of the paper is as follows. In section 2 we
set up the problem and solve the Laplace equation for the vacuum 
field outside the current sheet. This solution serves as a 
boundary condition to derive a closed equation for the potential and 
current in the space charge region using the continuity of the
electric field. In section 3  we specialize to the case of narrow
beams and make certain approximations.  We then solve the approximate
problem iteratively in section 4 
using a direct variational method.  The results are presented in
section 5.  We also consider  there the asymptotics of the current
density in  the limit $a\to 0$.  In section 6 we 
extend our approach to cylindrical current beams.  Section 7 is
devoted to a discussion of results.

\bigskip    
\noindent 2.{\it Formulation of problem}.

\smallskip
We will use dimensionless units: linear sizes are measured as 
fractions of the inter-electrode distance $D$, the potential  
$\phi(x,y)$ is given as a fraction of the inter-electrode voltage 
$V$, and the current density $j(x)$ is in the units of the
Child-Langmuir current density $j_0=V^{3/2}\sqrt{2e}/9\pi
D^2\sqrt{m}$, obtained in the limit $a=\infty$ [5]; $e$ and $m$ are
the electron charge and mass. 

\vskip-3cm
\hskip-0.6cm
\epsfig{file=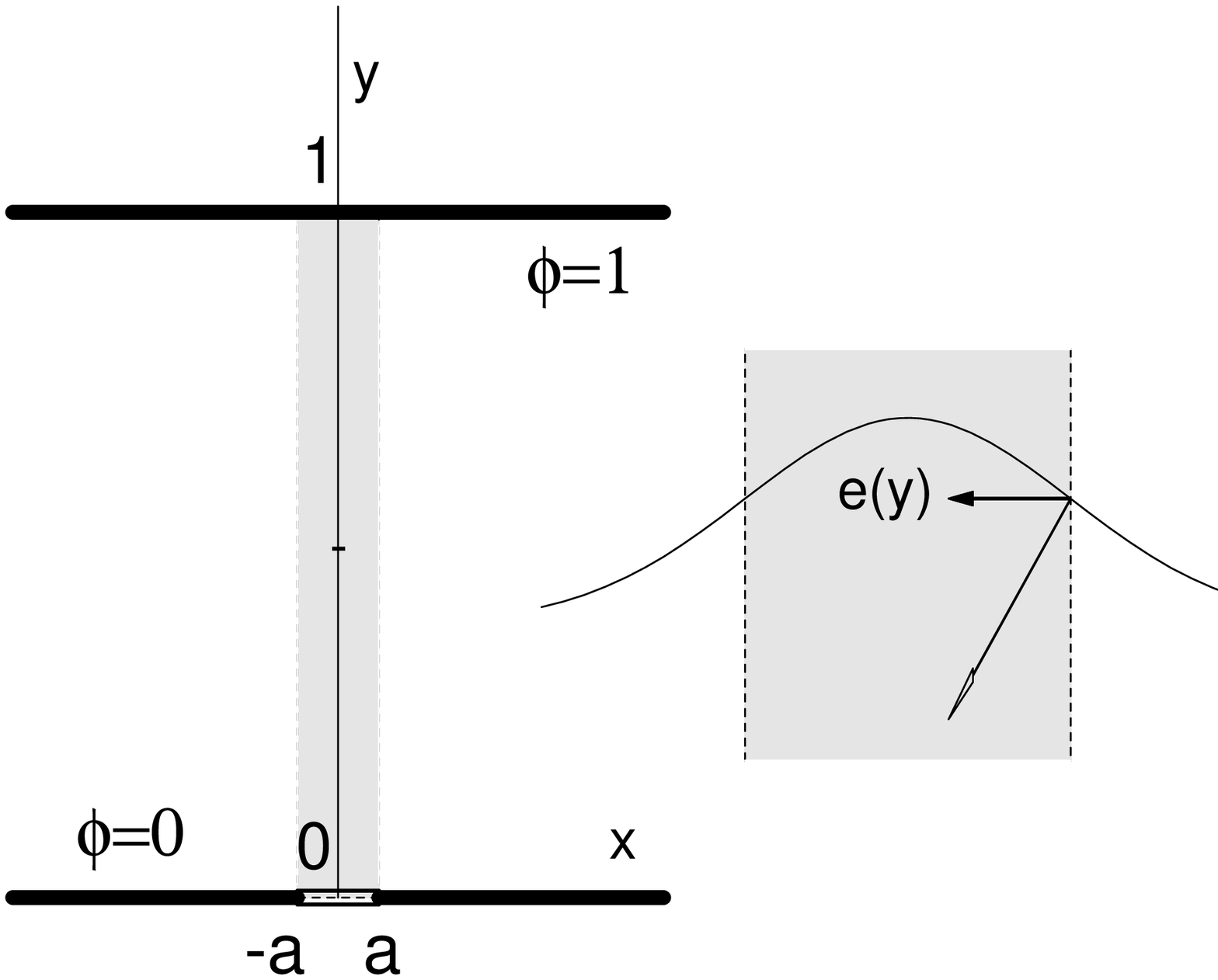, width=8cm,height=8cm}
\vskip-0.3cm
{\small FIG. 1. Cross-section of the current with an expanded
schematic graph of a contour of constant potential and electric field} 

\bigskip
The nonlinear Poisson equation which determines both the potential
$\phi(x,y)$ between the electrodes and the current density $j(x)$ for
a current sheet, see Fig.\ 1, 
has the form [9] $$
{\partial^2\phi\over \partial x^2}+{\partial^2\phi\over 
\partial y^2}={4j(x)\over 9 \sqrt{\phi(x,y)}}.\eqno(1)$$
Eq.\ (1) is to be solved
with the boundary conditions
$$
\phi(x,0)=0,\ \phi(x,1)=1. \eqno(2a)$$
The unknown space charge limited current $j(x)$, in the region 
$|x| < a$, is to be determined from the condition  
$$ {\partial\phi\over\partial y}(x,0)=0,
\ {\rm for}\ |x|<a.\eqno(2b)$$
For $|x| > a$, $j(x) = 0$ and (1) reduces to the Laplace equation in
vacuum.  We assume that (1) and (2) are well posed but leave a
mathematical investigation of this point for the future.

A natural approach to solving (1) is to separate it into a
standard boundary value problem for the Laplace equation outside the
space charge region, $|x|>a$, and a nonlinear inner problem in the 
rectangle, $-a\leq x\leq a,\ 0\leq y\leq 1$. Taking the 
potential $\psi(y)=\phi(a,y)$ on the sheet surface as given 
and using the symmetry $\phi(x,y)=\phi(-x,y)$ we find $\phi(x,y)$ for 
$x> a$.  From this one calculates the external electric field in the
$x$-direction $e_+(y)=-{\partial \phi\over \partial x}(a,y)$ 
at $x = a_+$.  The continuity of the potential and electric field at
the space charge boundary gives the additional  boundary conditions
$\psi(y)=\phi(a,y)$ and $e_-(y) = e_+(y) = e(y)$ needed
for solving (1) in the rectangle $|x| < a$, $0 \leq y
\leq 1$.  
\bigskip 

\noindent  {\it Vacuum Fields}

\medskip
We want to solve the Laplace equation (1) in the region $x>a$ with
the  boundary conditions, 
$$\phi(x,0)=0,\ \phi(x,1)=1,\ \phi(a,y)=\psi(y),\eqno(3)$$
and 
$$\phi(x,y) \to y \quad {\rm for} \quad  x \to \infty,$$
expecting $\psi(y)$ to be monotone with finite first derivatives.
Taking the potential in the form of a Fourier series
$$\phi(x,y)=y+{2\over \pi}\sum_{k=1}^\infty {a_k\over k} e^{\pi k(a-x)}
\sin {\pi ky},\ x>a\eqno(4)$$
(3) is satisfied if $$
a_k=\int_0^1{d\psi\over dy}(y)\cos{\pi k y}dy.\eqno(5)$$

An illustrative (and useful for our computations) example is the case 
$\psi(y)=y^{4/3}$.  Eq. (5) then gives $$
a_k={4\over 3}\Re\left [(-ik\pi)^{-4/3}\gamma\left ({4\over 3},-ik\pi
\right )\right ],$$
where $\gamma(u,v)$ is the incomplete gamma function [10].

Using (4) and (5) the $x$-component of the electric field at the beam 
boundary $x=a$ can be written as$$
e(y)=-{\partial\phi\over\partial x}(a,y)= 2\sum_{k=1}^\infty 
{a_k\sin {\pi ky}}.\eqno(6)$$
Substituting (5) into (6) and summing up the series explicitly we get 
an alternative form for $e(y)$ $$
e(y)=\sin {\pi y}\int_0^1 {\psi'(t)-\psi'(y)\over \cos{\pi t}-
\cos{\pi y}}dt.\eqno(7)$$
We note that Eq.(7) can also be derived by using conformal mapping, 
as in [9]. For $\psi(y)=y^2$ Eq.(7) becomes $$
e(y)={4\over\pi}\int_0^y\ln{\tan{\pi t\over 2}}dt.$$

\bigskip\noindent
{\it The inner region}

\medskip
Eq.\ (1) together with boundary conditions (2) and (7) now define the
space charge limited current problem in the rectangle $|x| < a$, 
$0 \leq y \leq 1$, the shaded region in Fig.\ 1.
Also shown there is an expanded cross section
together with a schematic drawing of an equipotential 
surface and electric field vector with its $x$-component.
As already noted the CLL potential ($a=\infty$) is $\phi_0(x,y)=
y^{4/3}$ and $j(x)=1$. For finite width of the current sheet 
$\phi(x,y)\to\phi_1(y)=y$ as $|x|\to\infty$, and we expect $\phi(x,y)$ inside
the beam, $|x|<a$, to satisfy the inequality $y^{4/3} \leq \phi(x,y) \leq y$.

\bigskip\noindent 3. {\it Narrow Beams}

\medskip
Integrating (1) between $x = 0$ and $x = a$, we obtain
$$-e(y) + {\partial^2 \over \partial y^2} \int_0^a \phi(x,y)dx = {4
\over 9} \int_0^a {j(x) \over {\sqrt \phi(x,y)}}dx. \eqno(8)$$
Eq.(8) can be rewritten as$$
-e(y) + a{d^2 \phi(x_1,y) \over dy^2}={4aj(x_2)\over 9\sqrt
{\phi(x_2,y)}},\eqno(9)$$
where $0\leq x_1(y),x_2(y)\leq a$.
Assuming now that for $a \ll 1$ there is only a small variation
in $\phi(x,y)$ for $0 \leq x \leq a$, at least when $y$ is not too
close to zero or one, we approximate (9) by setting $\phi(x_1,y) =
\phi(x_2,y)=\psi(y)$ and $j(x_2) = j$, where  $j$ is a parameter, 
close to the average current density, which will be determined  
together with $\phi(x,y)$. We expect that in 
the limit $a \to 0$, $e(y)$, $\psi(y)$ and $aj$ will become 
independent of $a$ (see section 6).  We therefore take the equation$$
-e(y) + a{d^2 \psi(y) \over dy^2}={4aj\over 9\sqrt{\psi(y)}}\eqno(10)$$
together with (7) as a suitable approximation for $a \ll 1$ and 
try to solve them numerically. The boundary conditions for $\psi$
come from (2): $\psi(0)=0,\ \psi'(0)=0,\ \psi(1)=1$.

An alternate way to arrive at (10), which also gives some information
about the beam profile, is 
to approximate $\phi(x,y)$ for $a \ll 1$ and $y$ not too close to zero
or one , as 
$$\phi(x,y) ~~ \simeq ~~ \psi(y)+{a \over 2}\left (1-{x^2 \over a^2}
\right ) e(y)\eqno(11)$$
so that $e(x,y)\simeq {x \over a} e(y).$
In the same spirit we would have
$$j(x)~~ \simeq ~~ j\left (\eta + \xi  
{x^2 \over a^2}\right ), \eqno(12)$$
where $j(\eta +\xi/3)$ is the mean current density and $\xi$ 
represents the relative height of the current wings (expecting 
$\eta$ to be close to $1$). Substituting (11), and (12) in (1)
we obtain two ordinary equations, namely (10) and$$
9{a^2-x^2\over 8aj}{d^2 e\over dy^2} \approx
{\eta + \xi x^2/a^2\over\sqrt{\psi+e(y)(a^2-x^2)/2a}}-
{1\over\sqrt{\psi}}.\eqno(13)$$
The ratio $$
\zeta (y)=- {e(y)a\over 2\psi(y)}\eqno(14)$$
gives the relative depth of the potential profile (11) at its deepest
point ($x=0$) as a function of $y$.

\bigskip
\noindent 4. {\it Numerical Algorithm}

\medskip
To solve (10) numerically we write $\psi$ in the form$$
\psi (y)=y^{4/3}+\sum_{m=1}^n{c_m f_m(y)},\eqno(15)$$
where the $f_m$ are a set of twice differentiable functions 
with the properties $f_m(0)=f'_m(0)=f_m(1)=0,\ m=1,2,...,n$
to satisfy the boundary condition (2) for $\psi(y)$.
 
The boundary electric field is similarly expressed as $$
e(y)=e_0(y)+\sum_{m=1}^n{c_m e_m(y)}.\eqno(16)$$
For computing $e_m(y),\ m=0,1,..,n$, we use (6)   
with the coefficients, given by (5), $$
a_k^m=\int_0^1 f'_m(y)\cos{\pi k y}dy.\eqno(17)$$
After this we apply the direct variational 
method to minimize the functional $\Phi(c_1,c_2,...,c_n;j)$ $$
\Phi =\int_{y_{min}}^{y_{max}}\left [e(y)
-a{d^2\psi(y)\over dy^2}+{4aj\over 9\sqrt{\psi(y)}}\right ]^2
y^{4/3}dy\eqno(18)$$
in terms of the parameters $c_m$ and $j$. The factor $y^{4/3}$ 
is used in (18) for regularization.  ``Neutralization'' of the 
nonlinearity of $1/\sqrt{\psi(y)}$ in (18) is achieved by an iteration
procedure where at each step we use
parameters $c_1,c_2,...,c_n$ evaluated 
in the previous step. The functional $\Phi$ thus becomes
bilinear in terms of its parameters and we solve repeatedly a linear
algebraic system$$
{\partial \Phi\over \partial j}=0,\ \ {\partial \Phi\over 
\partial c_m}=0,\ m=1,2,..,n\eqno(19)$$
until its solutions stabilize.

A serious problem in this variational computation is a good 
choice of the trial functions $f_m$.  We want in particular 
the right
behavior near the singular points $y=0$ and $y=1$. It is easy 
to see from (6) that $e(0)=e(1)=0$, but their
derivatives are generally divergent. On the other hand if the
error in a small neighborhood of these points is 
not large their overall impact on $e(y),\ 0<y<1$ is not
significant. To eliminate these regions in computing $\Phi$
we set $y_{min}\approx 0.09$ and $y_{max}\approx 0.99$ in (18).
We monitor the validity of our algorithm by two
indicators: (1) the minimum value of $\Phi$ after the iterations
get stabilized and (2) by deviations of the left part of Eq. (10)
multiplied by $9\sqrt{\psi(y)}/4a$ from the current $j$,
at $y=0.1,0.2,...,0.9$. For flexibility in approximating 
$\psi(y)$ and in order to keep the matrix 
of the system (19) away from any degeneracy we choose for $f_m(y),\ m=2,3,
...,10$ a set of partially overlapping functions of finite support.
The function $f_1(y)$ which corrects the CLL $f_0(y)$ for 
small $y$ is taken of the form $e^{-\beta y}y^{4/3}$ with $\beta 
\sim 40-60$.

To carry out our scheme practically we calculate in advance the 
partial electric fields
$e_m(y)$ on a homogeneous grid of  1000 points and then
apply the iterations.  These usually converge rapidly (after less 
than 10 steps) while each iteration takes a very short time. Some
series converge 
rather slowly (as $k^{-4/3}$), but even for a precision of $10^{-9}$,
i.e about $10^5$ terms in the series,
the computation of all partial fields is very fast.

After finding $\psi(y)$ and $e(y)$ the same, but much simpler, 
procedure can be applied for evaluating the parameters $\eta$ and 
$\xi$ in (12). We use for Eq.(13) the least squares method,  
where the weight function in the functional is chosen in such a
way  as to eliminate the derivatives of $e(y)$, and the computation
is done around the middle of the beam.

\bigskip
\noindent 5. {\it Results for the current sheet}.

\medskip
In Table 1 we present the results of the variational procedure
described in the last section for different values of $a$.  We show 
there the parameter $j$ which is found by (19) and 
the mean deviation $\Delta$ of the corresponding quantity, defined
in terms of $\psi(y)$ and $e(y)$ in Eq.(10), on the segment 
$0.1\leq y\leq 0.9$. The precision of
computation is controlled by the ratio $\Phi/\Phi_0$ at the stationary 
point, where $$
\Phi_0=\left ({4aj\over 9}\right )^2\int {y^{4/3}\over \psi(y)}dy.$$

$$\left |{\matrix{a&0.2&0.1&0.05&0.02&0.01&0\cr
j*a&0.549&0.447&0.394&0.360&0.347&0.326\cr
\Delta&0.074&0.080&0.075&0.062&0.053&0.041\cr
\Phi/\Phi_0&.0052&.0061&.0055&.0043&.0044&.0067\cr
\xi&0.042&0.032&0.021&0.009&0.004&~\cr
\delta&0.026&0.021&0.014&0.006&0.003&~\cr}}\right |$$
\centerline {\small TABLE 1. Flat current sheet}

\bigskip
Also shown there are the depth $\xi$ of the 
current density dip across the beam and the quantity$$
\delta=1-\eta -{\xi\over 3},$$
which gives the relative difference between the value $j$ computed
by solving Eq.(10) and the mean current density $\bar j(x)$.
The last column in Table 1 exhibits the limiting value of these
quantities when $a \to 0$ using a procedure described below.

\bigskip
\noindent {\it The limit $a \to 0$}.

\medskip
Our model assumes that the Larmor radii of electrons do not
exceed  $a$ and this makes the limit $a \to 0$ physically questionable
since this would require the magnetic field to become unreasonably 
strong. Nevertheless the limit $a\to 0$
is very interesting both mathematically and practically for 
evaluation of the prefactor in the scaling law for $j$.

Looking at Eq.\ (9) we expect that 
in the limit $a \to 0$ the term $aj(x_2)\to \lambda$, independent of
$a$, $\psi(x_2,y) \to \psi(y)$, and $a\frac{d^2 \phi(x,y)}{dy^2} \to
0$ except at $y=0,y=1$. The limiting $\psi(y)$ and $e(y)$ will then 
satisfy the equation 
$$
e(y) + {4\lambda \over 9\sqrt{\psi(y)}} = 0 \eqno(20)$$
with boundary conditions which we do not know.  We therefore solved
(20) 
in a truncated interval $y_1 = 0.09 \leq y \leq 0.998 = y_2$ using
the same routine as before.  The result of the computation is
presented in the last column of Table 1.  We believe now that by
improving and extending the set of trial functions $f_n(y)$ one can
gradually take $y_1$, $y_2$ closer to 0 and 1 respectively,
but we think that $\lambda \simeq 0.3$ is a good approximation. 

To get an ``exact'' equation for $\psi$ and $\lambda$ we integrate
Eq. (20) over an interval $[y_1,y_2]$ $0 < y_1 < y_2 < 1$ and 
sum the series in (5) and (6) to  yield a nonlinear 
eigenvalue problem $$
{9 \over 4\pi} \int_0^1 \psi^\prime(t) \ln \left | {\cos \pi y_1 - cos \pi
t \over \cos \pi y_2 - \cos \pi t} \right | dt = \lambda
\int_{y_1}^{y_2} {dt\over \sqrt{\psi(t)}}, \eqno(21)$$
where the boundary conditions are not needed. We postpone
investigation of (21) for the future.

\bigskip
\noindent 6. {\it Cylindrical beam}

\medskip
We consider now the more common case of electron beams of 
a compact cross section 
when the emitting part of the cathode as well as the cross section of 
the beam are circular. In cylindrical coordinates Eq. 1 takes the form$$
{1\over r}{\partial \over\partial r}\left (r{\partial\phi \over
\partial r}\right )+{\partial^2\phi\over 
\partial y^2}={4j(r)\over 9\sqrt{\phi(r,y)}},\eqno(22)$$
with the boundary conditions 
$$\phi(r,0)=0,\ \phi(r,1)=1;\ {\partial\phi\over\partial y}(r,0)=0,
\ {\rm for}\ r<a\eqno(23)$$
and $j(r)=0$ when $r>a$.  Fig.1 represents now the beam cross section 
which passes through its axis $r=0$.  

Carrying out again a Fourier expansion for $\phi(r,y)$ in the vacuum 
region $r>a$ yields the potential as a series$$
\phi(r,y)=y+{2\over \pi}\sum_{k=1}^\infty {a_k\over k} {K_0(\pi kr)
\over K_0(\pi ka)}\sin {\pi ky}, r \geq a \eqno(24)$$
where $a_k$ can be found by (5) with $\psi(y)=\phi(r,y)$ at $r=a$. 
Calling now $e(y)$  the radial component of the electric field, at
$r=a$, leads to the analogue of (6) $$
e(y)=-{\partial\phi\over\partial r}(a,y)=
2\sum_{k=1}^\infty {{K_1(\pi ka)\over K_0(\pi ka)}a_k\sin {\pi ky}}.
\eqno(25)$$
Here and in (24) $K_0,K_1$ are the modified Bessel functions which 
decay exponentially at infinity.

We can apply the same technique as in (9)-(12) and obtain instead of 
Eq.(10)$$
-2e(y)+a{d^2\psi\over d y^2}={4ja\over 9\sqrt{\psi(y)}}.\eqno(26)$$
Eqs.(13) and (19) for the current and potential profiles are the same. 
The factor $2$ in (26) shows the stronger effect of the surrounding
electric fields on the cylindrical beam compared 
with that on the narrow sheet. The same numerical scheme
as in part 4 yields the results shown in Table 2

\vskip0.2cm
$$\left |{\matrix{a&0.2&0.1&0.05&0.02&0.01\cr
j*a&5.574&5.532&5.452&5.422&5.410\cr
\Delta&0.090&0.069&0.052&0.037&0.036\cr
\Phi/\Phi_0&.0051&.0029&.0023&.0026&.0029\cr
\xi&.1089&.0566&.0293&.0119&.0059\cr}}\right |$$
\centerline {\small TABLE 2. Cylindrical beam}
\vskip0.3cm
\noindent

The current density behaves like $j\sim 5.4/a$ and therefore the total 
beam current becomes
proportional to $a$, when $a\ll 1$. The prefactor for cylindrical beams 
is significantly higher than the one for current sheets because
in this case the electron flow is completely surrounded by the vacuum 
field, but a compact expression like (21) for the limiting parameter 
$\lambda$ and the boundary potential $\psi$ is impossible here.
The profiles of the current density and potential are 
flatter than in the current sheet.

\bigskip \noindent 7.  {\it Discussion}

\medskip
As already noted the current density $j$ scales approximately as
$a^{-1}$ and thus will be large when the beam is 
narrow with the total current  of the sheet becoming 
independent of the sheet thickness (provided by the unlimited
emissivity of the cathode). The current density
grows slightly near the beam boundaries, but even for the flat sheet
whose thickness is $0.4$, which is not small, this rise is less than
$10\%$ (the evaluation of this quantity is more reliable for
$a< 0.1$). On the other hand when $y$ is small the potential across 
the beam width varies much stronger. In Fig.2 we show the electric 
field $e(y)$ for $a=0.1$ and the maximum deviation $\zeta (y)$ (14) 
of the equipotential surface from the horizontal for $a=0.1$ and 
$a=0.02$.

When $a\to 0$ the potential $\psi(y)$ approaches the solution of
(21). This function does not differ too much from  $y^{4/3}$. Only 
when $y$ is small does
$\psi(y)$ increase substantially. It has also an irregular behavior
near $y=1$ where $\psi''(y)\approx 4j/9$, since $e(1)=0$. On the other
hand $\psi''(y)$ is of order of $1$ almost everywhere on the segment
$(0,1)$ because if $\psi''(y)>C,\ C\geq 0$ on an interval $y_1\leq
y\leq y_2$ then, using the inequality $\psi'(y)\geq 0$, we would get
$\psi(y_2)-\psi(y_1)>C(y_2-y_1)^2/2$ which, if $C$ is large, will
contradict the condition $\psi(y)\leq 1$. Therefore the term $e(y)$ is
dominant almost everywhere in the left side of (10) for $a \ll 1$. 
\vskip-3cm
\hskip -0.5cm 
\epsfig{file=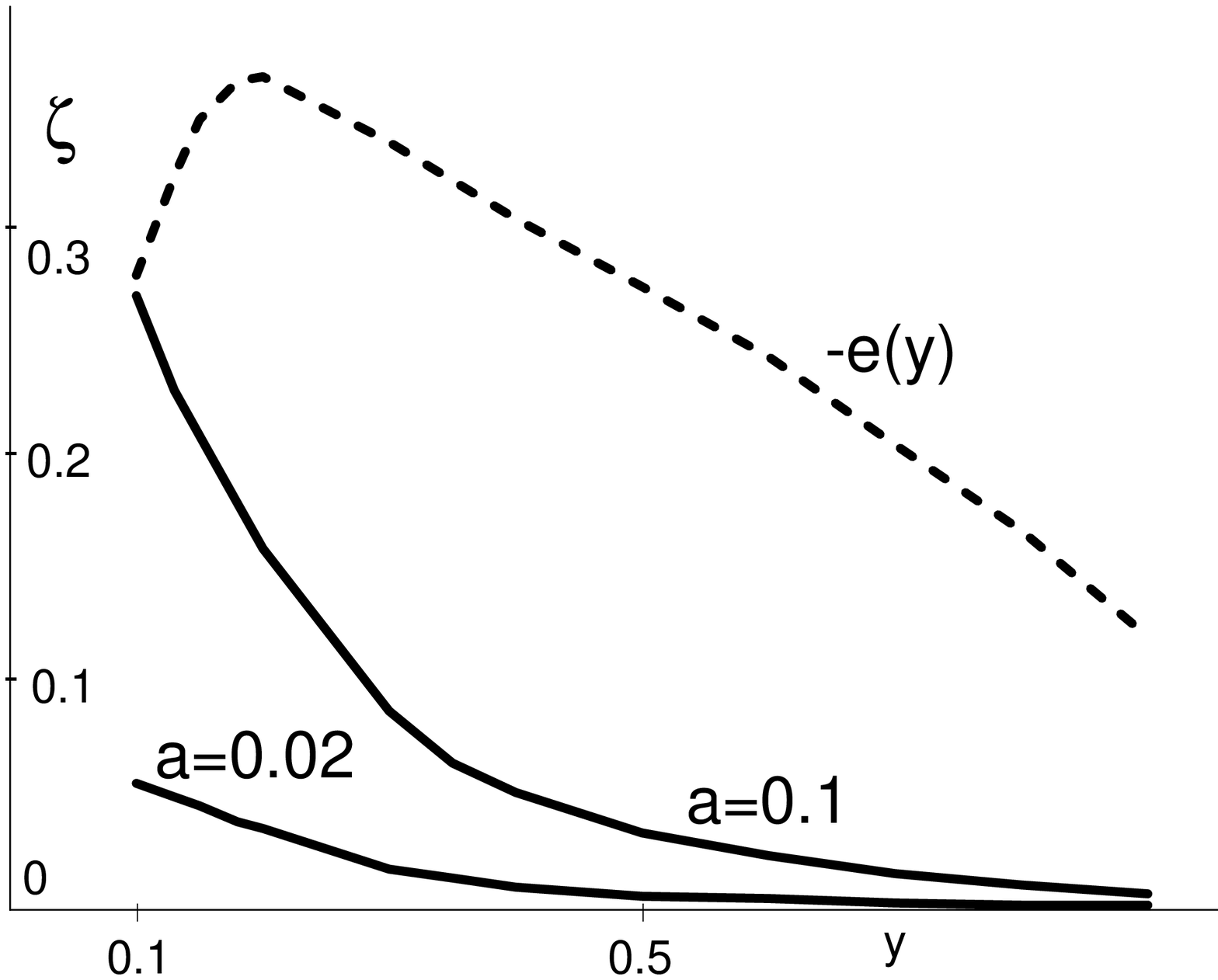, width=8cm,height=10cm}

{\small FIG. 2. Current sheet. Plots of $\zeta(y)$ for $a=0.1,\ 0.02$
(solid lines) and $e(y)$ for $a=0.1$ (dashed line)} 
\vskip 0.5cm
The equipotential surfaces deviate
from horizontal planes only when $y$ is small (in the case of
$a=0.01$ this effect is of lower order). Clearly our method of
computation, which neglects the $x$-derivatives higher than $2$, is
not accurate near $y=0$ where the electron density has a complicated
structure in the $x$-direction. We believe nevertheless (see discussion
in part 6) that these short intervals do not play a crucial role
because the current is limited by the whole space charge distributed
between the electrodes. 

There is a simple generalization of the parabolic shape assumption
used in (11),(12) for $\phi$ and $j(x)$ for larger $a$ [11]. This uses 
for the approximate solution of Eq.(1) the following ansatz$$
\phi(x,y)=\psi(y)+e(y)w(x),\ j(x)=j[\eta -\xi w(x)],\eqno(27)$$
where $w(x)=(\cosh{ga}-\cosh{gx})/k\sinh{ga}$ and the positive 
parameter $g$ can be taken from our previous work [9] ($g\approx
3.88$). Substituting (27) into (1) yields two ordinary differential 
equations similar to (10) and (13). The advantage of the scheme
used here for narrow beams is the transparent relationship between the 
exact Eq.(9) and our main Eq.(10) which is approximate for $a\not = 0$ 
but becomes exact, see (20), when $a\to 0$. 

\bigskip
{\it Acknowledgments}. Research supported by AFOSR Grant \# F49620-01-0154.
and by NSF Grant DMR 01-279-26.  We thank R. Barker for useful
discussions.


\vskip-0.6cm
\begin{thebibliography}{12}

\bibitem{[1]} A.S.Gilmour, Jr., {\sl Microwave Tubes} (Artech House, Dedham,
MA, 1986); P.T.Kirstein, G.S.Kino, and W.E.Waters, {\sl Space Charge
Flow} (McGraw-Hill, New York, 1967); A.Valfells, D.W.Feldman, M.Virgo, 
P.G.O'Shea, and Y.Y.Lau, Phys. Plasmas {\bf 9}, 2377 (2002).

\bibitem{[2]} J.W.Luginsland, Y.Y.Lau, R.J.Umstattd, and J.J.Watrous,
Phys. Plasmas {\bf 9}, 2371 (2002);  R.J.Umstattd and J.W.Luginsland, 
Phys. Rev. Lett. {\bf 87}, 145002 (2001)

\bibitem{[3]} J.W.Luginsland, Y.Y.Lau, and R.M.Gilgenbach,
Phys. Rev. Lett. {\bf 77}, 4668 (1996); Y.Y.Lau, Phys. Rev. Lett. 
{\bf 87}, 278301 (2001); Y.Y.Lau, P.J.Christenson, and D.Chernin, 
Physics of Fluids B{\bf 5}, 4486 (1993).

\bibitem{[4]}  R.J.Umstattd, D.A.Shiffler, C.A.Baca,
K.J.Hendricks, T.A.Spencer, and J.W.Luginsland, Proc. SPIE
Int. Soc. Opt. Eng. {\bf 4031}, 185 (2000); D.C.Barnes and R.A.Nebel, 
Phys. Plasmas {\bf 5}, 2498 (1998); R.A.Nebel and D.C.Barnes, Fusion 
Technology {\bf 38}, 28 (1998).

\bibitem{[5]} C.D.Child, Phys. Rev. {\bf 32}, 492 (1911); I.Langmuir,
Phys. Rev. {\bf 2}, 450 (1913); I.Langmuir and K.B.Blodgett, Phys. 
Rev. {\bf 22}, 347 (1923); I.Langmuir and K.B.Blodgett, Phys. Rev. 
{\bf 24}, 49 (1924).

\bibitem{[6]} J.R.Pierce, {\sl Theory and Design of Electron Beams},
(Van Nostrand, New York, 1954).

\bibitem{[7]} P.W.Hawkes and E.Kasper, {\sl Principles of Electron
Optics} Vol.2 (Academic Press, London, 1989), Chap.46.

\bibitem{[8]} J.Rouse et al., in {\sl Proceedings of SPIE} Vol.
3777, 65 (1999); M.A.Monastyrski, A.G.Murav'ev, and V.A.Tarasov,
in {\sl Proceedings of SPIE} Vol.4187, 2 (1999).

\bibitem{[9]} A.Rokhlenko and J.L.Lebowitz, Phys. Rev. Lett. 
{\bf 91}, 085002-(1-4) (2003).

\bibitem{[10]} M.Abramowitz and A.Stegun (editors), {\sl Handbook of
Mathematical Functions} (Wiley, New York, 1984).

\bibitem{[11]} A.Rokhlenko and J.L.Lebowitz, in preparation.

\end{thebibliography}
\end{document}